\documentclass[runningheads]{svmult}

\usepackage{graphicx}  
\usepackage{subeqnar}  
\usepackage{cropmark} 
\usepackage{physprbb}  

\newcommand{\greeksym}[1]{{\usefont{U}{psy}{m}{n}#1}}
\newcommand{\umu}{\mbox{\greeksym{m}}}

\begin{document}
\title*{The~Monte~Carlo~Event~Generator~DPMJET-III\footnote{SLAC-PUB-8740,
Talk given at the Conference ``Monte Carlo 2000'', Lisbon, Portugal, 
23-26 Oct. 2000}}
\toctitle{The Monte Carlo Event Generator DPMJET-III}
\titlerunning{DPMJET-III}
\author{Stefan Roesler\inst{1}
\and    Ralph Engel\inst{2}
\and    Johannes Ranft\inst{3} }
\authorrunning{Stefan Roesler et al.}
\institute{SLAC, P.O. Box 4349, Stanford CA 94309, USA
\and       University of Delaware, Bartol Res. Inst., Newark DE 19716, USA
\and       University of Siegen, D--57068 Siegen, Germany}
\maketitle              
\begin{abstract}
A new version of the Monte Carlo event generator \textsc{Dpmjet} is presented.
It is a code system based on the Dual Parton Model and unifies all features 
of the \textsc{Dtunuc-2}, \textsc{Dpmjet}-II and \textsc{Phojet}1.12 event 
generators.
\textsc{Dpmjet-III} allows the simulation of hadron-hadron, hadron-nucleus, 
nucleus-nucleus, photon-hadron, photon-photon and photon-nucleus interactions 
from a few GeV up to the highest cosmic ray energies.
\end{abstract}
%
\section{Introduction}
\vspace*{-0.1cm}
Hadronic collisions at high energies involve the production of
particles with low transverse momenta, the so-called \textit{soft}
multiparticle production. The theoretical tools available at present
are not sufficient to understand this feature from QCD and
phenomenological models are typically applied instead. The
Dual Parton Model (DPM) \cite{dpm-a} is such a model and its fundamental
ideas are presently the basis of many of the Monte Carlo (MC)
implementations of soft interactions in codes used for Radiation
Physics simulations.

Many of these implementations are however limited in their application by,
for example, the collision energy range which they are able to describe
or by the collision partners (hadrons, nuclei, photons) which the model can
be used for. With respect to modern \textit{multi-purpose} codes for particle
interaction and transport these limitations at high energy are clearly often a
disadvantage.

In this paper we present the \textsc{Dpmjet}-III code system, a MC event 
generator based on the DPM which is unique in its wide range of application.
\textsc{Dpmjet}-III is capable of simulating hadron-hadron, hadron-nucleus,
nucleus-nucleus, photon-hadron, photon-photon and photon-nucleus interactions
from a few GeV up to the highest cosmic ray energies. 

In the present paper we give an overview over the different components and
models of \textsc{Dpmjet}-III and present a few examples for comparisons of 
model results with experimental data.
\vspace*{-0.1cm}
\section{The Concept of the Program}
\vspace*{-0.1cm}
\textsc{Dpmjet}-III is the result of merging all features of the 
event generators \textsc{Dpmjet}-II \cite{dpmjet2-a,dpmjet2-b} and 
\textsc{Dtunuc-2} \cite{dtunuc2-a,dtunuc2-b} into one single code system.
The latter two codes are similar in their underlying concepts, however they
differ in the Monte Carlo realization of these concepts, in particular,
of the DPM.

Whereas individual nucleon-nucleon collisions in \textsc{Dpmjet}-II are
simulated based on the \textsc{Dtujet} model \cite{dtujet}, 
\textsc{Dtunuc-2} is using \textsc{Phojet}1.12 \cite{phojet-a,phojet-b}.
Since \textsc{Phojet} describes not only hadron-hadron interactions but also
hadronic interactions involving photons, \textsc{Dtunuc-2} allows also
the simulation of photoproduction off nuclei. Therefore, the strength of
\textsc{Dtunuc-2} is in the description of photoproduction and nuclear
collisions up to TeV-energies. On the other hand, \textsc{Dpmjet}-II is 
widely used to simulate cosmic-ray interactions up to 
the highest observed energies \cite{dpmjet2-b}.

However, many program modules in \textsc{Dpmjet}-II and \textsc{Dtunuc-2} 
are also identical. Examples are the Glauber-Gribov formalism for
the calculation of nuclear cross sections \cite{diagen}, the formation-zone 
intranuclear cascade \cite{fzic}, the treatment of excited 
nuclei \cite{fzic-a,fzic-b} and
the \textsc{Hadrin}-model for the description of interactions below 5 GeV
\cite{hadrin}.

The core of \textsc{Dpmjet}-III consists of \textsc{Dtunuc-2} and
\textsc{Phojet}1.12. In addition all those features of \textsc{Dpmjet}-II 
were added which were not part of \textsc{Dtunuc-2} so far.
This includes, for example, quasi-elastic neutrino interactions 
\cite{qelnu} and certain baryon-stopping diagrams~\cite{barystop-b}.
\vspace*{-0.1cm}
\section{Models Implemented in DPMJET-III}
\vspace*{-0.1cm}
%
%
\vspace*{-0.1cm}
\subsection{The Realization of the Dual Parton Model}
\vspace*{-0.1cm}
The DPM combines predictions 
of the large $N_c,N_f$ expansion of QCD \cite{Veneziano74}
and assumptions of duality \cite{Chew78} with
Gribov's reggeon field theory \cite{Gribov67a-e}.
\textsc{Phojet}, being used for the simulation of 
elementary hadron-hadron, photon-hadron and 
photon-photon interactions
with energies greater than 5 GeV, implements the DPM as a two-component model
using Reggeon theory for soft and perturbative QCD for hard interactions.
In addition to the model features as described in detail in \cite{PhD-RE},
the version 1.12 incorporates a model for high-mass diffraction
dissociation including multiple jet production and recursive
insertions of enhanced pomeron graphs (triple-, loop- and double-pomeron
graphs). In the following only the new features are briefly discussed.

High-mass diffraction dissociation is simulated as pomeron-hadron or
pomeron-pomeron scattering, including multiple soft and hard
interactions \cite{Bopp98a}. To account for the nature of the pomeron being a
quasi-particle, the CKMT pomeron structure function \cite{Capella96a}
with a hard gluonic
component is used. These considerations refer to pomeron exchange
reactions with small pomeron-momentum transfer, $|t^2|$. For large
$|t^2|$ the rapidity gap production (e.g. jet-gap-jet events) is 
implemented on the basis the color evaporation model \cite{Eboli98a}.

Extrapolating the two-channel eikonal-unitarization of a hadron-hadron 
amplitude as used in \textsc{Phojet} to very high energies raises the
question of the treatment of enhanced graphs which become more and more
important at high energy and lead to large multiplicity fluctuations. 
A full amplitude calculation including enhanced graphs is very involved
and not suited for a Monte Carlo implementation. Therefore, based on the 
results of \cite{Kaidalov86b-e}, we use the simpler approach of 
interpreting each 
soft pomeron as the sum of a series of a bare soft pomeron and enhanced 
graphs (Froissaron). In practice, this results in the simulation of 
possibly recursive subdivisions of a single Froissaron cut into various 
other configurations such as, for example, two cut pomerons or 
a single cut pomeron and a diffractive scattering. However, the current
implementation should only be considered as a first step toward a
consistent treatment of enhanced graphs at very high energy 
because of its limitation to soft interactions. 
\vspace*{-0.1cm}
\subsection{Hadronic Interactions Involving Photons}
\vspace*{-0.1cm}
The photon is assumed to be a superposition of a \textit{bare} photon 
interacting in direct processes and a \textit{hadronic} photon interacting
in resolved processes.

The description of interactions of the hadronic photon with nuclei
is based on the Generalized Vector Dominance Model (GVDM)
\cite{GVDM}. Photons are assumed to fluctuate into quark-antiquark states $V$
of a certain mass $M$ and the interaction is described as scattering of
the hadronic fluctuation on the nucleus. Correspondingly, the scattering
amplitude $a_{VA}$ reads \cite{dtunuc2-a}
\begin{eqnarray}
a_{VA}(s,Q^2,M^2,\vec{B})=\int\prod_{j=1}^{A}d^3r_j\
\psi_A^\star\
a_{VA}(s,Q^2,M^2,\vec{B}_1,\ldots,\vec{B}_A)\
\psi_A  \\
a_{VA}(s,Q^2,M^2,\vec{B}_1,\ldots,\vec{B}_A)=
\frac{i}{2}\left(1-\prod_{\nu=1}^{A} 
\left[1+2ia_{VN}(s,Q^2,M^2,\vec{B}_\nu)\right]\right)
\end{eqnarray}
where $a_{VA}$ is expressed in terms of interactions on individual
nucleons $N$ according to the Gribov-Glauber picture (see below).
The model is limited to low photon-virtualities $Q^2$ satisfying the
relation $Q^2\ll 2m_N\nu$ ($\nu$ and $m_N$ being the photon energy and
nucleon mass).
For individual $q\bar{q}$-nucleon interactions it is sufficient to
consider only two generic $q\bar{q}$-states, the first one grouping
$\rho^0$, $\omega$ and $\phi$ and $\pi^+\pi^-$-states up to the $\phi$-mass
together and the second one including all $q\bar{q}$-states with higher
masses~\cite{phojet-a}. 

Direct photon interactions are treated as either gluon-Compton scattering
or photon-gluon fusion processes on a single nucleon. The consideration
of so-called anomalous interactions allows a steady transition between
direct and resolved interactions \cite{dtunuc2-a}.

Finally, an interface to \textsc{Lepto}6.5 \cite{lepto6} allows to
simulate deep-inelastic scattering off nuclei.
\vspace*{-0.1cm}
\subsection{The Gribov-Glauber Multiple Scattering Formalism}
\vspace*{-0.1cm}
The Monte Carlo realization of the Gribov-Glauber multiple scattering
formalism follows the algorithms of \cite{diagen} and allows the 
calculation of total, elastic, quasi-elastic and production cross
sections for any high-energy nuclear collision. 
Parameters entering the hadron-nucleon scattering amplitude (total cross
section and slope) are calculated within \textsc{Phojet}.

For photon-projectiles ideas of the GVDM have been incorporated
in order to correctly treat the mass of the hadronic fluctuation and its 
coherence length as well as pointlike photon interactions \cite{dtunuc2-a}.
Realistic nuclear densities and radii are used for light nuclei
and Woods-Saxon densities otherwise.

During the simulation of an inelastic collision the above formalism samples
the number of ``wounded'' nucleons, the impact parameter of the collision and
the interaction configurations of the wounded nucleons. Individual 
hadron(photon,nucleon)-nucleon interactions are then described by 
\textsc{Phojet} including multiple hard and soft pomeron exchanges, initial
and final state radiation as well as diffraction. 

As a new feature, \textsc{Dpmjet}-III allows the simulation of
enhanced graph cuts in non-diffractive inelastic hadron-nucleus and
nucleus-nucleus interactions. For example, in an event with two wounded
nucleons, the first nucleon might take part in a non-diffractive
interaction whereas the second one scatters diffracively producing only
very few secondaries. Such graphs are prediced by the Gribov-Glauber
theory of nuclear scattering but are usually neglected.

Finally, all color neutral
strings are hadronized according to the Lund model as implemented in
\textsc{Pythia} \cite{pythia6-a,pythia6-b}.
\vspace*{-0.1cm}
\subsection{The Intranuclear Cascade and Break-up of Excited Nuclei}
\vspace*{-0.1cm}
The treatment of intranuclear cascades in spectator prefragments and their
subsequent fragmentation is largely identical to the one described in Refs.
\cite{fzic-a,fzic-b}. 

Particles created in string fragmentation processes are
followed on straight trajectories in space and time. A certain formation time
is required before newly created particles can re-interact in the spectator
nuclei. These re-interactions are of low energy and are described by 
\textsc{Hadrin}\cite{hadrin} based on parameterized exclusive interaction channels. In 
nucleus-nucleus collisions the intranuclear cascade is calculated in both the 
projectile and target spectators.

Excitation energies of prefragments are calculated by summing up the
recoil momenta transfered to the respective prefragment by the hadrons leaving 
the nuclear potential (a constant average potential is assumed). 
The prefragments are
assumed to be in an equilibrium state and excitation energy is dissipated
by the evaporation of nucleons and light nuclei and by the emission of photons.
%
\vspace*{-0.1cm}
\section{Comparison to Experimental Data}
\vspace*{-0.1cm}
Since \textsc{Dpmjet}-III is the result of merging \textsc{Dpmjet}-II and
\textsc{Dtunuc-2} its predictions have to be in agreement to experimental
data where there was agreement for the two latter codes before. However, this
has to be proven again.
Here, only a few examples are given which should represent the large
amount of comparisons of \textsc{Dpmjet}-III results with experimental data
which exist.

\begin{figure}[htbp]
\setlength{\unitlength}{1cm}
\begin{picture}(16,4.8)
   \put(-0.8,-0.8){\scalebox{0.75}
   {\includegraphics*[bb=0 0 1027 802,height=18.5cm]{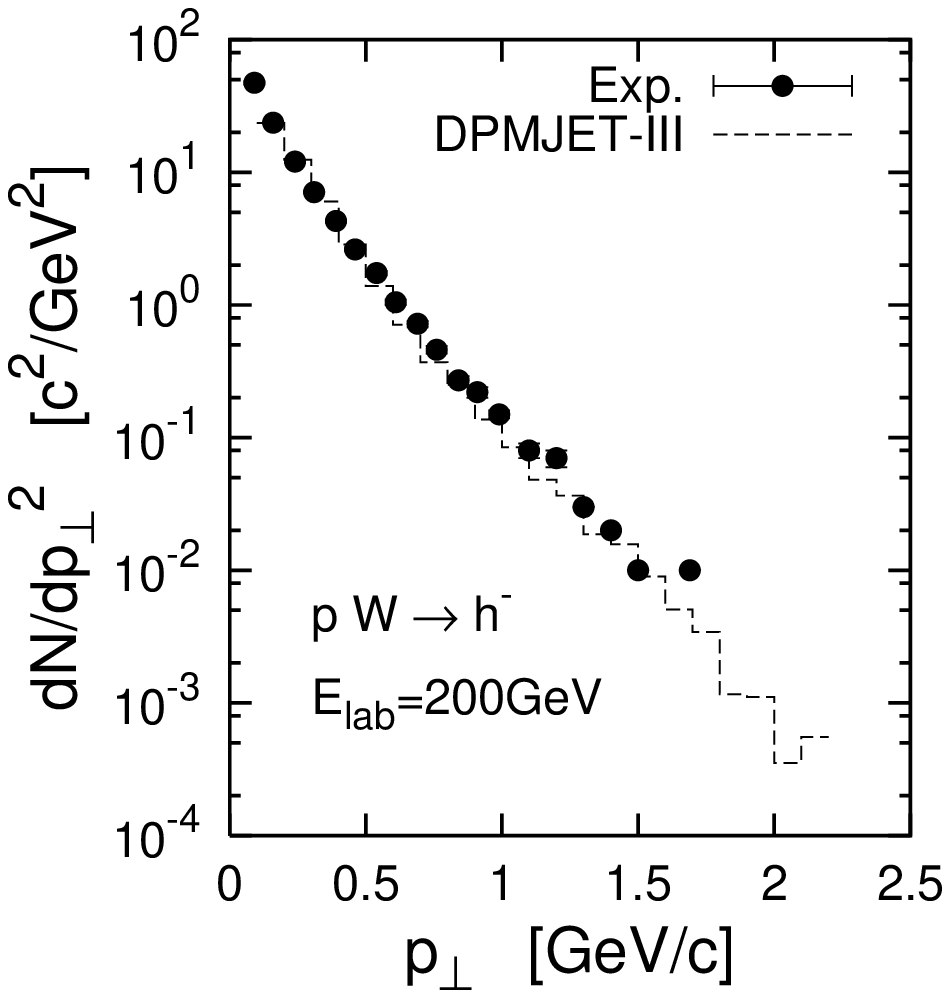}}
                  }
   \put(5.2,-0.8){\scalebox{0.75}
   {\includegraphics*[bb=0 0 1027 802,height=18.5cm]{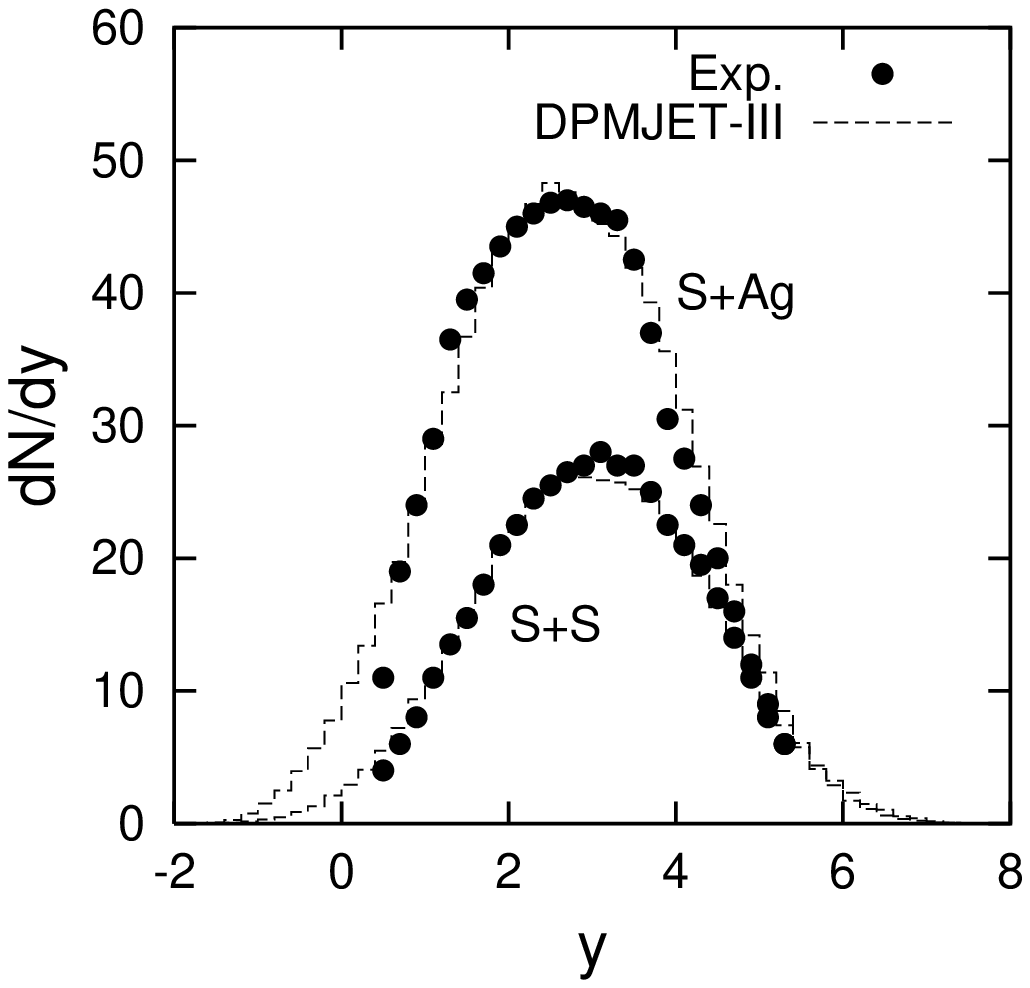}}
                  }
   \put(0.5,0.0){ {(\textbf{a})} }
   \put(6.0,0.0){ {(\textbf{b})} }
\end{picture}
\caption{
Negatively charged hadron production in nuclear collisions at 200 GeV/nucleon
}
\label{comp1}
\vspace{-0.45cm}
\end{figure}

\begin{figure}[htbp]
\setlength{\unitlength}{1cm}
\begin{picture}(16,4.3)
   \put(-0.5,-0.8){\scalebox{0.75}
   {\includegraphics*[bb=0 0 1027 802,height=17.5cm]{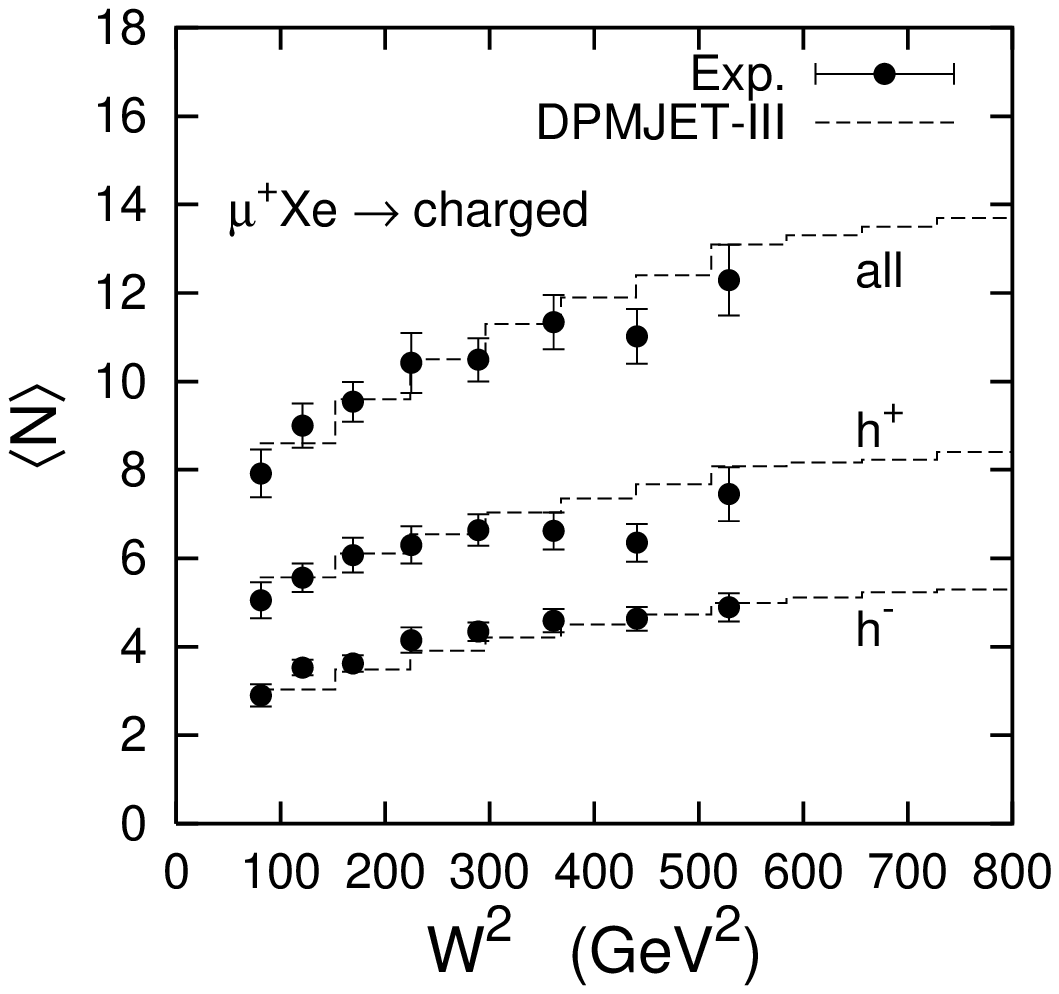}}
                  }
   \put(5.0,-0.8){\scalebox{0.75}
   {\includegraphics*[bb=0 0 1027 802,height=18.5cm]{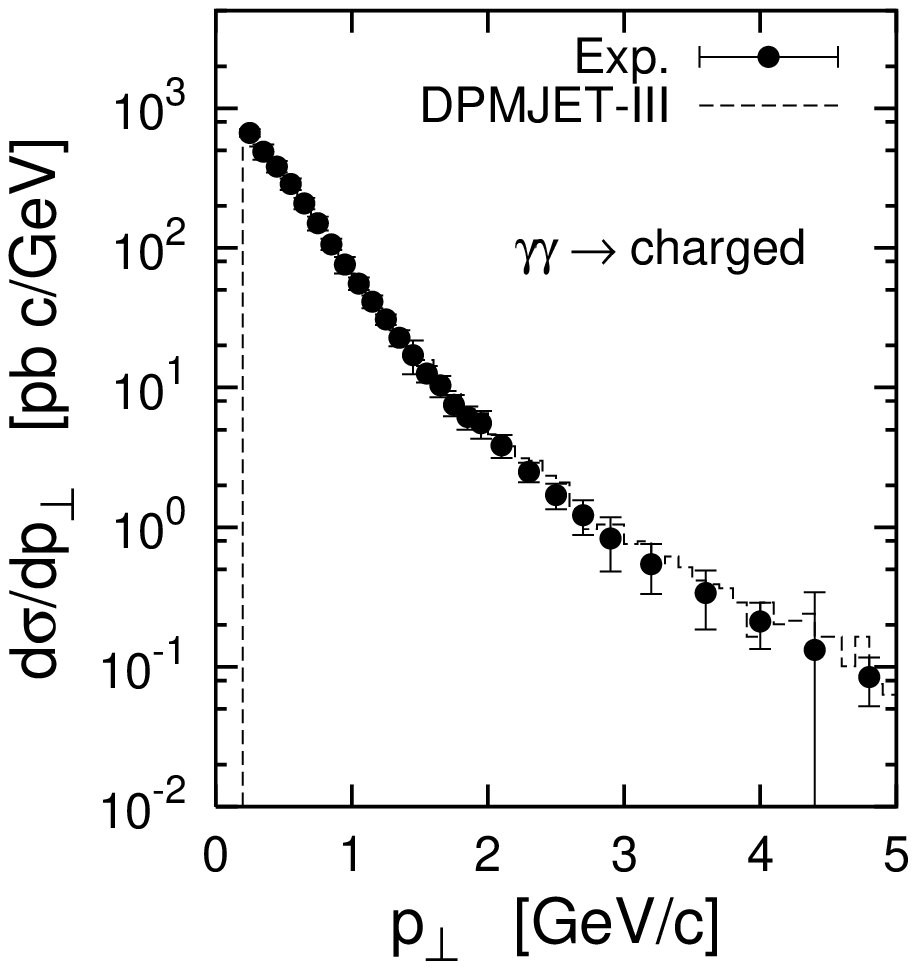}}
                  }
   \put(0.5,0.0){ {(\textbf{a})} }
   \put(6.0,0.0){ {(\textbf{b})} }
\end{picture}
\caption{
Comparison of \textsc{Dpmjet}-III results to data on interactions involving
photons.
}
\label{comp2}
\end{figure}

Fig. \ref{comp1}a shows the transverse momentum distribution of negative
hadrons from p-W collisions together with data \cite{helios}. 
The rapidity distributions of negative hadrons in central S-S and S-Ag
collisions are compared to data \cite{na35} in Fig. \ref{comp1}b.
%
Two examples for interactions involving photons are given in Fig. \ref{comp2}.
Hadronic interactions of muons are described by the radiation off the muon
of a quasi-real photon and the subsequent interaction of the photon.
Fig. \ref{comp2}a shows
average multiplicities of charged hadrons from $\mu$-Xe interactions at 490 GeV
compared to data \cite{e665}.
In Fig. \ref{comp2}b the calculated inclusive transverse momentum cross 
section of charged particles produced in two-photon collisions at LEP is
compared to the combined data set of the ALEPH, L3, and OPAL Collaborations 
for low-$Q^2$ deep inelastic scattering \cite{Finch99a}.
\vspace*{-0.9cm}
%
\vspace*{-0.1cm}
\section{Conclusions}
\vspace*{-0.1cm}
A new version of the \textsc{Dpmjet} event generator is presented. 
\textsc{Dpmjet}-III is based on \textsc{Dpmjet}-II, \textsc{Dtunuc-2} and 
\textsc{Phojet}1.12 and unifies all features of these three event generators 
in one single code system. It has been thoroughly tested and, in due time, will
largely superseed the older \textsc{Dpmjet} and \textsc{Dtunuc} versions.

It is presently not advisable to use the code for very low-energy 
nucleus-nucleus collisions (below $\approx 10-20$ GeV). This requires 
further testing and tuning of parameters.
Furthermore deficiencies exist in the description of some effects observed 
in heavy ion collisions at AGS- and SPS-energies (e.g.\ strangeness 
enhancement, transverse energy flow).

The code is available on request from the authors (Stefan.Roesler@cern.ch,
Johannes.Ranft@cern.ch) and further information can be found on the World
Wide Web (\verb?http://home.cern.ch/sroesler/dpmjet3.html?).
\vspace*{-0.1cm}
\section{Acknowledgements}
\vspace*{-0.1cm}
The work of S.R. and R.E. is supported by the Department of Energy under 
contracts DE-AC03-76SF00515 and DE-FG02-91ER40626, respectively.

\end{document}